\documentstyle[aps,prl,preprint,epsfig]{revtex}
\tightenlines

\begin{document}
\draft
\title{{\it Ab Initio} Theoretical Description of the
       Interrelation between Magnetocrystalline Anisotropy and
       Atomic Short-Range Order}

\author{S.S.A. Razee$^a$, J.B. Staunton$^a$, B. Ginatempo$^b$,
        F.J. Pinski$^c$, and E. Bruno$^b$}
\address{ $^a$Department of Physics, University of Warwick,
          Coventry CV4 7AL, United Kingdom}
\address{$^b$Dipartimento di Fisica and Unit\'a INFM,
         Universit\'a di Messina Salita Sperone 31,
         I-98166 Messina,
         Italy}
\address{$^c$Department of Physics, University of Cincinnati,
         Ohio 45221, USA}

\date{\today}

\maketitle

\begin{abstract}
The cubic lattice symmetry of ferromagnetic homogeneously
disordered alloys is broken when a compositional modulation is
imposed. This can have a profound influence on the
magnetocrystalline anisotropy energy (MAE). We describe our
{\it ab initio} theory of this effect and use the framework of
concentration waves with the electronic structure described
within the spin-polarised relativistic Korringa-Kohn-Rostoker
coherent-potential approximation. We find that ordering
produces a 2 order of magnitude increase in the MAE as well
as altering the equilibrium direction of magnetisation.
Using the same theoretical framework we also examine
directional compositional order produced by magnetic
annealing with an explicit study of permalloy. \\
\end{abstract}

\pacs{PACS numbers: 75.30.Gw, 75.50.Cc, 75.60.Nt, 75.50.Ss}

Magnetocrystalline anisotropy (MCA) of ferromagnetic materials
containing transition metals has become the subject of intensive
theoretical and experimental research because of the
technological implications for high-density magneto-optical
storage media \cite{lmf,tw}. Potential materials for these
applications need to exhibit a substantial perpendicular
magnetic anisotropy (PMA), which is evidently due to an
intrinsic magnetic anisotropy in the crystal lattice
strong enough to overcome the extrinsic macroscopic
shape anisotropy favoring an in-plane magnetisation. Whereas
in ultrathin films and multilayers the PMA is due to surface
\cite{had} and interface \cite{ghod,pb} effects respectively,
in thick films of transition metal alloys it is an
intrinsically bulk magnetic property which leads
to PMA. Such systems are particularly interesting for
magneto-optic recording because in addition to PMA they exhibit
large Kerr effect signals as well as being chemically stable and
easy to manufacture. There has been much effort directed
towards an understanding of the mechanism of MCA from a
first-principles electronic structure point of view to aid
future magnetic material design but since MCA arises from
spin-orbit coupling, which is essentially a relativistic
effect \cite{hjfj} this means that a fully relativistic
electronic structure framework is desirable.

Empirically compositional structure is found to have a
profound influence on both the magnitude of the
magnetocrystalline anisotropy energy (MAE) as well as the
equilibrium or {\it easy} magnetisation direction.
Compositional order lowers the lattice symmetry of the
homogeneously disordered alloy and this enhances
its MAE. For example, the measured MAE ($ \sim 130 \mu$eV) of
ordered CoPt alloy (CuAu or $L1_0$ type) \cite{gh} is some
two orders of magnitude larger than that of its disordered
counterpart ($ \sim$3 $ \mu$eV), and with a different easy
axis. In this letter we provide the first quantitative
calculations of the effect of compositional order on MAE
via a study of CoPt.

We also note that thick fcc-Co$_c$Pt$_{1-c}$ \cite{dw,mm,wg}
and fcc-Co$_c$Pd$_{1-c}$ \cite{dw} films exhibit PMA of a size
comparable with that of multilayers. Such a large PMA is quite
unexpected for systems which should have effectively bulk
cubic structure. Now whereas tensile strain and large negative
magnetostriction coefficients may account for some of this in
the Co$_c$Pd$_{1-c}$ films, it cannot for the thick
Co$_c$Pt$_{1-c}$ ones \cite{sh}. Rather recent experiments have
confirmed that although Co$_c$Pt$_{1-c}$ films seem to be nearly
homogeneously disordered there are actually more Co-Co nearest
neighbors in-plane and very few out-of-plane \cite{mm,tat,pwr},
i.e. there is presence of some atomic short-range order (ASRO).
The existence of this compositional order produces
{\it internal interfaces} analogous to Co/Pt multilayers and
this has been suggested as a likely cause of the strong PMA in
fcc-Co$_c$Pt$_{1-c}$ films \cite{wg,tat}. This suggestion is
strengthened by the fact that films grown at higher temperatures,
in which enhanced bulk diffusion tends to destroy the in-plane
compositional order, show no PMA. Very recently Kamp {\it et al}
\cite{pk} have shown from magnetic circular x-ray dichroism
measurements that chemical ordering is also responsible for the
enhanced MAE in thick Fe$_{0.5}$Pd$_{0.5}$ films. These
observations clearly underline a correlation between MCA
and ASRO. In this letter we provide, for the first time, a
detailed theoretical analysis based on first-principles
calculations which shows how chemical order significantly
influences MCA and compare the MAE of ordered and disordered
CoPt alloys. Moreover, we use the same approach to model
other hitherto unfabricated structures. This is pertinent now
that it is possible to {\it tailor} compositionally modulated
films to obtain better magneto-optic recording
characteristics \cite{rc}.

Upon ordering CoPt undergoes a modest tetragonal lattice
distortion ($c/a=0.98$) which also lowers the symmetry. From
magnetostriction data however we can estimate this to contribute
only 15$\%$ of the MAE. The measured magnetostriction
coefficient, $\lambda$, of pure fcc-Co ($5 \times 10^{-5}$) is
of the same size as that of CoPt (-$4 \times 10^{-5}$)
although with different sign. From `first-principles'
electronic structure calculations Wu {\it et al} \cite{rqw}
have shown that the rate of change of the MAE with lattice
strain is proportional to $\lambda$ and thus in fcc-Co and
CoPt should be roughly of the same magnitude. Thus from
Wu {\it et al}'s \cite{rqw} calculation of the MAE of Co for a
range of $c/a$ ratios we estimate that a 2\% tetragonalization
will change the MAE only by about 20$\mu$eV. Consequently we
conclude that it is the compositional order that is primarily
responsible for the large MAE of CoPt.

In previous work, we presented a theory of MCA of disordered
bulk cubic alloys \cite{our1,our2} within the framework of
spin-polarised relativistic Korringa-Kohn-Rostoker
coherent-potential approximation (SPR-KKR-CPA) \cite{he}. In
this letter we set up a new framework to investigate the
effects of compositional order, both short and long-ranged.
Any compositionally modulated alloy can be specified by
site-dependent concentrations $ \{ c_i \} $ which themselves
can be written as a superposition of static concentration
waves (CWs) \cite{agk}, i.e.,
\[
c_i = c + \frac{1}{2} \sum_{\bf q} \left[
           c_{\bf q} e^{i {\bf q} \cdot {\bf R}_i } +
           c_{\bf q}^\ast e^{-i {\bf q}
           \cdot {\bf R}_i } \right],
\]
with wave-vectors {\bf q} and amplitudes $ c_{\bf q} $. Usually
only a few CWs are needed to describe a
particular ordered structure. For example, the CuAu-like
$ L1_0 $ structure (Fig. 1) is set up by a single
CW with $ c_{\bf q}=\frac{1}{2} $ and
$ {\bf q}=(001) $, and the [111]-layered CuPt-like $ L1_1 $
structure is set up by a CW with
$ c_{\bf q}=\frac{1}{2} $ and
$ {\bf q}=( \frac{1}{2} \frac{1}{2} \frac{1}{2} ) $
({\bf q} is in units of $ \frac{2 \pi}{a}$, $ a $ being the
lattice parameter).

The grand-potential for the interacting electrons in
an alloy with composition $ \{ c_i \} $ and magnetised along
the direction $ {\bf e}_1 $ at a finite temperature
$ T$ is given by \cite{blg,jbs},
\begin{eqnarray*}
\Omega (\{ c_i \}; {\bf e}_1 ) = \nu Z & - &
       \int_{- \infty}^\infty
d \varepsilon f(\varepsilon , \nu ) N (\{ c_i \}, \varepsilon ;
{\bf e}_1 ) \\
& + & \Omega_{DC} (\{ c_i \}; {\bf e}_1 ),
\end{eqnarray*}
where, $ \nu $ is the chemical potential, $ Z $ the total valence
charge, $ f(\varepsilon , \nu ) $ the Fermi factor,
$ N (\{ c_i \}, \varepsilon ; {\bf e}_1 ) $ the integrated
electronic density of states, and
$ \Omega_{DC} (\{ c_i \}; {\bf e}_1 ) $
the `double-counting' correction. The MAE of
the inhomogeneous alloy can be characterised by the difference
\[
K(\{ c_i \}) = \Omega (\{ c_i \}; {\bf e}_1 ) -
               \Omega (\{ c_i \}; {\bf e}_2 ).
\]
Assuming that $ \Omega_{DC} (\{ c_i \}; {\bf e}) $
is generally unaffected by the change in the magnetisation
direction, we get,
\begin{eqnarray*}
K(\{ c_i \}) & = & \int_{- \infty}^\infty d \varepsilon
   f(\varepsilon , \nu_1 ) \left[
   N (\{ c_i \}, \varepsilon ; {\bf e}_1 ) -
   N (\{ c_i \}, \varepsilon ; {\bf e}_2 ) \right] \\
   & + & O ( \nu_1 - \nu_2 )^2.
\end{eqnarray*}
Note that the correction due to the change in the chemical
potential (from $ \nu_1 $ to $ \nu_2 $) with the magnetisation
direction is of second order in $ ( \nu_1 - \nu_2 ) $, and can be
shown to be very small compared to the first term \cite{our1}.
We now expand $ K(\{ c_i \}) $ around $ K_{CPA} (c) $, the MAE
of the homogeneously disordered alloy $A_c B_{1-c}$,
\begin{eqnarray}
K(\{ c_i \}) & = & K_{CPA} (c) + \sum_j \left.
\frac{ \partial K(\{ c_i \}) }{ \partial c_j }
\right|_{ \{ c_i = c \} } \delta c_j \nonumber \\
 & + & \frac{1}{2} \sum_{j,k} \left.
\frac{ \partial^2 K(\{ c_i \}) }{ \partial c_j \partial c_k }
\right|_{ \{ c_i = c \} } \delta c_j
\delta c_k + O ( \delta c )^3. \label{eq:k1}
\end{eqnarray}
The second term in Eq. (\ref{eq:k1}) vanishes if the number of
$ A $ and $ B $ atoms in the alloy is to be conserved. Now taking
the Fourier transform, we get the MAE of the compositionally
modulated alloy with wave-vector {\bf q},
\begin{equation}
K({\bf q}) = K_{CPA} (c) + \frac{1}{2}
\vert c_{\bf q} \vert ^2  \left[ S^{(2)} ({\bf q}; {\bf e}_1) -
S^{(2)} ({\bf q}; {\bf e}_2) \right], \label{eq:k2}
\end{equation}
and so, for example, the MAE of the CuAu-type $L1_0$ ordered
alloy is obtained by choosing $ {\bf q}=(001) $ and
$ c_{\bf q}=\frac{1}{2} $. Here
$ S^{(2)} ({\bf q}; {\bf e}) $ are the Fourier transforms of
the so-called direct correlation functions \cite{blg} which
determine the ASRO parameter $ \alpha ({\bf q}) $ in the
disordered phase,
$ \alpha ({\bf q}) = \beta c (1-c) / [ 1 - \beta c (1-c)
S^{(2)} ({\bf q}) ]$, ($ \beta = 1/k_B T $, $ k_B $ being the
Boltzmann constant). These have been calculated for many alloys,
both non-magnetic and ferromagnetic, in which up to now
relativistic effects were largely ignored and were compared
with diffuse X-ray and neutron scattering data \cite{blg,jbs}.
$ S^{(2)} ({\bf q}) $ is given by,
\begin{eqnarray*}
S^{(2)} ({\bf q}) & = & - \frac{Im}{\pi} \int_{- \infty}^\infty
  d \varepsilon f(\varepsilon , \nu ) \sum_{L_1 L_2 L_3 L_4}
  ( X^A - X^B )_{ L_1 L_2 } \\
 & & \times I_{ L_2 L_3 ; L_4 L_1 } ( {\bf q} )
  \Lambda_{ L_3 L_4 } ({\bf q} ),
\end{eqnarray*}
where,
\begin{eqnarray*}
\lefteqn{\Lambda_{ L_1 L_2 } ( {\bf q}) =
( X^A - X^B )_{ L_1 L_2} } \hspace{0.4in} \\
 & & - \sum_{L_3 L_4 L_5 L_6} X^A_{ L_1 L_5}
 I_{ L_5 L_3 ; L_4 L_6 } ({\bf q})
X^B_{ L_6 L_2} \Lambda_{ L_3 L_4 } ({\bf q} ),
\end{eqnarray*}
\begin{eqnarray}
I_{ L_5 L_3 ; L_4 L_6 } ({\bf q}) =
  \frac{1}{V_{BZ}} \int & d {\bf k} &
  \tau_{ L_5 L_3 } ( {\bf k} + {\bf q} )
  \tau_{ L_4 L_6 } ( {\bf k}) \nonumber \\
& - & \tau_{ L_5 L_3 }^{00} \tau_{ L_4 L_6 }^{00}, \label{eq:iq}
\end{eqnarray}
and $ X^{A(B)}=[(t_{A(B)}^{-1}-t_c^{-1})^{-1}+\tau^{00} ]^{-1}$.
Here, $ \tau^{00} $ is the site-diagonal path-operator matrix,
$ t_{A(B)} $ and $ t_c $ are the t-matrices
for electronic scattering from sites
occupied by $ A (B)$ atoms and the CPA effective potentials
respectively,
$ \tau ( {\bf k}) = [ t_c^{-1} - g ({\bf k}) ]^{-1} $,
and $ g ({\bf k}) $ is the KKR structure constants
matrix \cite{jsf}. The
spinodal ordering temperature below which the alloy orders into a
structure characterised by concentration wave-vector
${\bf q}_{max}$ is given by \cite{blg,jbs},
$ T_c = c(1-c) S^{(2)} ({\bf q}_{max} ; {\bf e})/k_B $, where
${\bf q}_{max}$ is the value at which
$S^{(2)} ({\bf q}; {\bf e})$ is maximal.

As with all MAE calculations, given the sizes of the energies
involved, numerical computation of $ K ({\bf q}) $ needs to be
done very carefully. The energy integration is done using a
complex contour described elsewhere \cite{our1}. The
Brillouin zone (BZ) integration is done using the adaptive
grid method \cite{eb} which has been found to be very
efficient and accurate. In this method one can preset the
level of accuracy of the integration by supplying an error
parameter $ \epsilon $. The integration in Eq. (\ref{eq:iq}) is
done with $ \epsilon = 10^{-6} $ which means that
$ S^{(2)} ({\bf q}; {\bf e}) $ which are of the order of
0.1 eV are accurate up to 0.1 $ \mu $eV. To achieve such level of
accuracy, we had to sample a large number of $ {\bf k}$-points in
the BZ. Also, owing to the form of the integrand in
Eq. ({\ref{eq:iq}) the integration has to be done using the full
BZ. Typically, in our calculations, we needed around
35,000 $ {\bf k}$-points for points on the energy contour 0.5 Ry
above the real axis. When the energy was 0.0001 Ry above the real
axis (and that is the closest point to the real axis, we need) we
required as many as 3 million
$ {\bf k}$-points for the same level
of accuracy. Furthermore, we have calculated
$ S^{(2)} ({\bf q}; {\bf e}_1) $ and
$ S^{(2)} ({\bf q}; {\bf e}_2) $
simultaneously ensuring that they
are calculated on the same grid, and thus cancelling out the
systematic errors if any. Therefore, we claim that the values of
$ K ({\bf q}) $ are accurate to within 0.1 $ \mu $eV.

We summarise the results for fcc-Co$_{0.5}$Pt$_{0.5}$ alloy in
Tables \ref{table1} and \ref{table2}. We note that,
$ S^{(2)} ({\bf q}) $ is maximum for the $ L1_0 $ structure,
implying that the alloy would order into this structure at 1360 K
as it is cooled from high temperature in  good agreement with
experiment (ordering temperature of 1000 K \cite{gh}). From
Table \ref{table2} we note that for $ {\bf q}=(001) $ and
$ {\bf q}=(100) $ which correspond to CuAu-like $ L1_0 $
ordered structure, with Co and Pt layers stacked along the
[001] and [100] directions respectively, the direction of
spontaneous magnetisation is along the [001] and [100] directions
respectively in excellent agreement with experiment. Also, the
MAE (58.6 $\mu $eV) is comparable to the experimental
value ($\sim 130 \mu $eV \cite{gh}).

In order to probe the relationship between the compositional
structure and MAE further, we also performed
calculations for some hypothetical structures. These are also
summarised in Table \ref{table2}. The case of
$ {\bf q}=( \frac{1}{2} \frac{1}{2} \frac{1}{2} ) $
corresponding to the CuPt-type $ L1_1 $ ordered structure with
Co and Pt layers perpendicular to the [111] direction of the
crystal produces a spontaneous magnetisation along the [111]
direction of the crystal. This may be close to the structure of
the thick [111] oriented disordered CoPt alloy films exhibiting
PMA \cite{dw,mm,wg} which is attributed to the existence of
internal interfaces \cite{tat}, analogous to Co and Pt layers
along the [111] direction. Our result is clearly consistent with
these observations. The point is that the magnetic anisotropy
(152$\mu$eV) intrinsic to this structure is nearly 3 times
larger than that of $ L1_0 $ structure. Indeed, we predict
that a [111]-oriented film will exhibit a markedly stronger
PMA than that of a [100]-oriented film.

The structure set up by CWs with
$ {\bf q}=(1 0 \frac{1}{2}) $ and $ (0 1 \frac{1}{2}) $
is also a [001]-oriented layered structure, but the layers are
not alternately pure Co and Pt planes, rather they are layers of
ordered Co and Pt (Fig. 1). Even in this case, we note that the
magnetisation is perpendicular to the layered structure, and
the magnitude of MAE is large. Similarly, for
$ {\bf q}=(\frac{1}{2} 0 1) $ and
$ (\frac{1}{2} 1 0) $ where the planes are stacked along the
[100] direction the magnetisation is also along the
[100] direction.

Evidently, the spontaneous magnetisation always tends to align
itself perpendicular to any layering in the structure.
In addition, the magnitude of MAE depends strongly on the
symmetry of the system, i.e. it increases when the symmetry is
lowered. The cubic symmetry of the homogeneously disordered alloy
quenches the orbital magnetic moment. In the $ L1_0 $ ordered
alloy the symmetry is lower and in the $ L1_1 $ layered structure
it is lower still, thus increasing the MAE in each case. On
detailed examination of the electronic structure of the
disordered alloy around the Fermi energy we find that the large
values of $ K ({\bf q}) $ near the BZ boundary arise from
van Hove singularities of the Bloch spectral function \cite{jsf}.
The number, location, and occupation of these depend on the
magnetisation direction and produce a large contribution
to the difference in the convolution
integrals (Eqs. (\ref{eq:k2}) and (\ref{eq:iq})).

Our theory can also be used to produce the first quantitative
description of the phenomenon of magnetic annealing. Here a soft
magnetic material develops directional chemical order when
annealed in a magnetic field \cite{sc}. We demonstrate this
effect for Ni$_{0.75}$Fe$_{0.25}$ (permalloy) and
Table \ref{table3} is a summary of the results. We calculate
$S^{(2)} ({\bf q},{\bf e})$ for permalloy in an applied
magnetic field of strength 600 Oe (same as used in the
experiment \cite{sc}) orientated along
${\bf e}=$ [001], [111], and [100] directions of the crystal. We
find that when the magnetic field is along [001] (column 2)
$ S^{(2)} ({\bf q}) $ is maximum for {\bf q}=(001) confirming
that ordering is favored along the direction of applied field.
Similar is the case when the magnetic field is along [100].
However, when the applied field is along [111] (column 6)
all the three orderings, namely, (100), (010), and (001)
are favored equally. Thus, in this case, we will get a
CuAu$_3$-type $ L1_2 $ ordering. The calculated transition
temperature 721 K is in good agreement with the experimental
value of 820 K \cite{jo}. Noting that the measured intensity
in a scattering experiment is proportional to the ASRO
parameter $ \alpha ( {\bf q} ) $, we estimate that for an
alloy cooled in a magnetic field along the [001] direction
the superlattice spot at {\bf q}=(001) will be 20\% more
intense than that at {\bf q}=(100) at a temperature 1 K
above the transition temperature.

In conclusion, we have presented a fully relativistic electronic
structure scheme to study the MCA of alloys and its dependence
upon compositional structure. We applied this theory to
fcc-Co$_{0.5}$Pt$_{0.5}$ and found that when the system is cooled
it tends to order into $L1_0$ layered-ordered structure and that
the spontaneous magnetisation tends to align itself perpendicular
to the layer stacking. These observations are in complete accord
with experiment. We also found that
if the layers are stacked along
the [111] direction then the MCA becomes larger which may be the
case in [111]-textured films. Within the same framework we have
also been able to explain the appearance of directional order in
Ni$_{0.75}$Fe$_{0.25}$ when it is annealed in a magnetic field.

We thank B.L. Gyorffy for many helpful discussions. This research
is supported by the Engineering and Physical Sciences Research
Council (UK), National Science Foundation (USA), and the TMR
Network on ``Electronic structure calculation of materials
properties and processes for industry and basic sciences''.

\begin{table}
\caption{Direct correlation function
$ S^{(2)} ({\bf q}; [001])$ for
different {\bf q}-vectors for Co$_{0.5}$Pt$_{0.5}$ alloy.}
\begin{tabular}{dddddd}
        &            & $ S^{(2)} ({\bf q}; [001]) $  & $ T_c $ \\
{\bf q} & Structure  & (eV)                          &  (K)  \\
\tableline
(000)                             & Clustering & -1.51 &   -   \\
(001)                             & $ L1_0 $   &  0.47 & 1360  \\
(100)                             & $ L1_0 $   &  0.47 & 1360  \\
($ \frac{1}{2} \frac{1}{2} \frac{1}{2} $) & $L1_1 $ & 0.29 &   \\
($ 1 0 \frac{1}{2} $)             &            &  0.19 &       \\
\end{tabular}
\label{table1}
\end{table}

\begin{table}
\caption{Magnetocrystalline anisotropy energy
$ K ({\bf q}) $ for several compositionally modulated CoPt
alloys characterised by different {\bf q}-vectors (the
respective ordered structures are shown in Fig. 1). The
reference system has the magnetisation along the [001]
direction (i.e. $ {\bf e_1} = [001] $).}
\begin{tabular}{dddd}
 & $ {\bf e}_2 = [111] $ & $ {\bf e}_2 = [100] $ & \\
 {\bf q} & $ K({\bf q}) $ & $ K({\bf q}) $ & Easy Axis \\
 & ($ \mu $eV) & ($ \mu $eV) & \\
\tableline
(001)                              &  -58.6 & -105.6 & [001] \\
(100)                              &   39.6 &  105.9 & [100] \\
($\frac{1}{2} \frac{1}{2} \frac{1}{2}$) & 152.0 & 0.0 & [111] \\
($ 1 0 \frac{1}{2} $)              & -158.7 & -236.5 & [001] \\
($ \frac{1}{2} 0 1 $)              &   85.0 &  236.3 & [100] \\
\end{tabular}
\label{table2}
\end{table}

\begin{table}
\caption{Direct correlation function
$ S^{(2)} ({\bf q}; {\bf e})$ for
different {\bf q}-vectors for Ni$_{0.75}$Fe$_{0.25}$ alloy.}
\begin{tabular}{ddddddd}
 & \multicolumn{2}{c}{{\bf e}=[001]}
 & \multicolumn{2}{c}{{\bf e}=[100]}
 & \multicolumn{2}{c}{{\bf e}=[111]} \\
 {\bf q} & $ S^{(2)} ({\bf q})$ & $ T_c $
         & $ S^{(2)} ({\bf q})$ & $ T_c $
         & $ S^{(2)} ({\bf q})$ & $ T_c $ \\
 & (meV) & K & (meV) & K & (meV) & (K) \\
\tableline
(100)   &  330.880 & & 331.158 & 720.9 & 330.973 & 720.5  \\
(010)   &  330.880 & & 330.880 & & 330.973 & 720.5  \\
(001)   &  331.158 & 720.9 & 330.880 & & 330.973 & 720.5  \\
($ \frac{1}{2} \frac{1}{2} \frac{1}{2} $) & -99.291 &
                                          & -99.291 &
                                          & -99.475 &  \\
($ 1 0 \frac{1}{2} $) & 99.284 & & 99.096 & & 99.222 & \\
($ \frac{1}{2} 0 1 $) & 99.096 & & 99.285 & & 99.222 & \\
\end{tabular}
\label{table3}
\end{table}

\begin{figure}
\caption{Some ordered structures and their representative
wave-vectors. {\bf q}=(001) generates the CuAu-type structure
with ordering along the [001] direction.
{\bf q}=($\frac{1}{2} \frac{1}{2} \frac{1}{2} $) generates
the CuPt-type structure. {\bf q}=($ 1 0 \frac{1}{2}$)
generates a layered structure with planes of an ordered
structure of $ A $ and $ B $ atoms stacked along the [001]
direction.}
\end{figure}

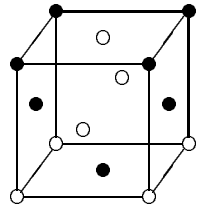
\end{document}